\documentclass[12pt]{article}

\usepackage{graphics}
\usepackage{amsmath}
\def\ni{\noindent}
\def\ph{{\phantom{...}}}
\def\={\phantom{..} = \phantom{..}}

\def\+{\phantom{..} + \phantom{..}}
\def\>{\phantom{..} > \phantom{..}}
\def\<{\phantom{..} < \phantom{..}}
\def\-{\phantom{..} - \phantom{..}}

\def\bq{\begin{quote}}
\def\eq{\end{quote}}
\def\be{\begin{equation}}
\def\ee{\end{equation}}
\def\bar{\begin{eqnarray}}
\def\ear{\end{eqnarray}}
\def\no{\nonumber}

\def\qp{quantum physics}

\def\prob{probability}
\def\probs{probabilities}
\def\ea{{\em et. al}}

\def\Sch{Schr{\"o}dinger}

\def\Schs{Schr{\"o}dinger's}

\def\Copism{Copenhagenism}

\def\Copists{Copenhagenists}

\def\vN{von Neumann}

\def\Poin{Poincar{\'e}}
\def\Ham{Hamiltonian}
\def\wf{wavefunction}

\def\cH{{\bf{H}}}

\def\sjN{\sum_{j=1}^N}
\def\sijN{\sum_{i,j=1}^N}
\def\skN{\sum_{k=1}^N}

\def\srmo{{\sqrt{ - 1}}}
\def\half{{\frac{1}{2}}}

\title{\bf On Non-Linear Quantum Mechanics\\[1in] and the Measurement Problem\\[1in] III. \Poin\ Probability and ... Chaos?\\[2in]}

\author{W. David Wick\footnote{email: wdavid.wick@gmail.com}}

\begin{document}
\maketitle
\pagebreak

\section*{Abstract}
Paper I of this series introduced a nonlinear version of quantum mechanics that blocks cats, 
and paper II postulated a random part of the wavefunction to explain outcomes in experiments such as Stern-Gerlach or EPRB, \cite{wick}. 
However, an {\em ad hoc} extra parameter was assumed for the randomness. Here I provide some analytic and simulation evidence that the nonlinear theory 
exhibits sensitive dependence on initial conditions in measurement scenarios, perhaps implying that the magnitude of randomness required 
is determined by structural features of the model, and does not require a free parameter.
\vfill\pagebreak

\section{Introduction}
Probability entered into quantum physics in a different manner than for the other sciences, including the other branches of physics.
Thirty years before Heisenberg and \Sch\ published, Henri \Poin\ pointed out in a book, \cite{poin}, that we tend to resort to \probs\ in situations where a small uncertainty in some component of
a system is amplified by the dynamics. In these situations, a ``universal" \prob\ law may result, and we do not require additional parameters to describe the underlying uncertainty.
Although \Poin\ discussed mostly physicist's models such as a mass on a spring, consider roulette. We assume that, provided the wheel is moving at an adequate rate, the croupier will be unable to 
coordinate her toss in such a way as to bias the outcome in a desired direction, and thus we adopt the \prob\ model of all outcomes equally likely. 
Modeling the croupier's hand motion isn't interesting or necessary.

By contrast, \prob\ appeared in \qp, in a paper by Max Born in 1926, \cite{born}, in a rather {\em ad hoc} fashion. Born noted that certain wavefunction components, when (the modulus) is squared, are positive and add to one,
and declared that they represent {\em \probs}. In a discussion of the unit sphere centered at the origin as part of a geometry problem, 
one would certainly note that $x^2 + y^2 + z^2 = 1$, but would not then propose
that these squared variables represent ``the probability of {\em finding} one of the them." Nevertheless, this interpretation was adopted by the Copenhagen school of quantum theorists. Then, in 1932,
John \vN\ axiomatized the idea, \cite{vnbook}, asserting that any self-adjoint operator on the quantum Hilbert space represented ``an observable", 
despite the fact that there are infinitely many such operators that can be constructed and for the vast majority no one knows of an apparatus that would measure it. 
(E.g., what device would measure $q^5\,p^{17}\,q^5$?). No attempt was made by the \Copists\ to explicate the random element justifying the appeal to \probs; they merely asserted that, 
if an operator was ever measured,
the outcomes would be eigenvalues, with probabilities given by the squared-modulus of the corresponding Hilbert space eigenvector. \Sch\ never accepted this interpretation of his \wf.

In paper II of this series, in order to explain random outcomes of experiments within \Schs\ paradigm (``\wf\ only"), I adopted a random component of the \wf, with a corresponding free parameter.
But every additional parameter requires dedicated experiments to measure it, and makes the whole theory harder to test. However, in the first paper I proposed a nonlinear extension of \Schs\ equation
which is \Ham\ and blocks cat formation through energy conservation. High-dimensional, nonlinear dynamical systems are well-known to exhibit some form of ``chaos", 
meaning sensitive dependence on initial conditions and seemingly unpredictable, ``random" orbits. 
This raises the possibility that the nonlinear quantum mechanics (NLQM) proposed in paper I exhibits such phenomena, which may imply that the magnitude of randomness
required to explain certain experiments (such as Stern-Gerlach or EPRB) is dynamically determined, and moreover it may be unnecessary to parameterize the unstable element (as for the croupier's hand).

Chaos is usually defined by exponential divergence of any two distinct initial states evolving over time. 
Call it ``strong chaos". Such extreme events may not be needed to explain outcomes in quantum experiments. Rather,
it would suffice for sensitive dependence on initial conditions to appear in certain ``crisis" situations created by the measurement scenario. Call it ``weak chaos". Here I provide some analytical
and simulation results indicating that such phenomena occur in the presence of the nonlinear energy postulated in paper I. Unfortunately, I was limited to very small systems (ten or fewer ``qubits")
due to computing limitations and the exponential increase of dimensionality ($q$ qubits implies a Hilbert space of dimension $2^q$ 
and a real dynamical system of dimension $2\,2^q$, e.g., for $q = 10$, 1024 and 2048 dimensions, respectively). And, to see anything interesting for such a small system 
necessitates an absurdly-large coupling constant in front of the nonlinear energy.
So my evidence for chaos is itself weak, but still suggestive of trends as well as illustrating many of the points made in papers I and II.

In section two I introduce finite-dimensional quantum mechanics and discuss the eigenvalues appearing in the linear case. In section three a nonlinear model 
is set up and the same issue discussed. In section four a measurement-like case is introduced and the computational and simulation results presented.
A Discussion section raises many issues for future research. Technical issues are relegated to the Computational Appendix.

\section{Expanding and Contracting Directions, Linear Case}

Throughout the paper I will be analyzing/simulating a finite-dimensional quantum system which, in the conventional linear case, takes the form:

\be
\srmo\,\frac{\partial \psi_i}{\partial t} \= \sjN \, H_{i,j}\,\psi_j.\label{scheq}
\ee

Here I have set $\hbar = 1$ and $\psi$ is a complex-valued vector of length $N$. 
(It is more usual to define Hilbert space vectors by the values of some operators, e.g., 
to write $\psi(s_1,s_2,...)$ where the ``spins" take values $\pm J$. I will revert to this notation
in a later section, but clearly this is just a matter of indexing.) 
$H$ is a complex-valued, $N$x$N$ matrix which is self-adjoint:

\be
H_{j,i}^{*} \= H_{i,j}.
\ee

\ni The asterisk denotes complex conjugate. 
The self-adjointness condition is fulfilled if:

\be
H_{i,j} \= K_{i,j} \+ \srmo\,L_{i,j},
\ee

\ni where $K$ and $L$ are real matrices symmetric and anti-symmetric respectively:

\be
K_{i,j} \= K_{j,i}; \ph L_{i,j} \= - L_{j,i}.
\ee

Equation (\ref{scheq}) defines a complex or ``quantum" dynamical system.
But I wish to consider (\ref{scheq}) as a real dynamical system (RDS). So let

\be
\psi_i \= Q_i \+ \srmo\,P_i,
\ee

\ni where $Q$ and $P$ are real $N$-vectors. (These are NOT the physical position and momentum, but rather should be thought of as the P's and Q's of a course in classical mechanics.)
Separating real and imaginary parts of (\ref{scheq}) then yields:

\bar
\no \frac{\partial Q_i}{\partial t} &\=& \sjN\,\left\{ \, L_{i,j}\,Q_j \+ K_{i,j}\,P_j\,\right\};\\
 \frac{\partial P_i}{\partial t} &\=& \sjN\,\left\{ \, L_{i,j}\,P_j \- K_{i,j}\,Q_j\,\right\}. \label{QPsys}
\ear

\ni It is easily checked that these are Hamilton's equations:

\bar
\no \frac{\partial Q_i}{\partial t} &\=& \frac{\partial}{\partial P_i}\, \cH;\\
 \frac{\partial P_i}{\partial t} &\=& \- \frac{\partial}{\partial Q_i}\, \cH,
\ear

\ni for the Hamiltonian (energy) function:

\be
\cH \= \sijN\,\left\{\,\half\,Q_i\,K_{i,j}\,Q_j \+ \half\,P_i\,K_{i,j}\,P_j + P_i\,L_{i,j}\,Q_j\,\right\}.
\ee

We will also be interested in the linearized or Jacobian dynamical system (JDS) associated to an RDS, defined by:

\bar
\no \frac{\partial \xi_i}{\partial t} &\=& \ph\ph\,\, \sjN\,\left\{\, \frac{\partial^2 \cH}{\partial P_i\,\partial Q_j}\,\xi_j \+ \frac{\partial^2 \cH}{\partial P_i\, \partial P_j}\,\eta_j \,\right\}\\
 \frac{\partial \eta_i}{\partial t} &\=& \- \sjN\,\left\{\, \frac{\partial^2 \cH}{\partial Q_i\,\partial Q_j}\,\xi_j \+ \frac{\partial^2 \cH}{\partial Q_i\, \partial P_j}\,\eta_j \,\right\} \label{Jacsystem}
\ear

\ni Here $\xi$ and $\eta$ are real $N$-vectors. The JDS represents the motion of a system which approximates the original for a small time interval, and hence is useful to understand local trends.
(Think of $\xi$ and $\eta$ evolving for a small time, with the P's and Q's held fixed.) For linear QM, the JDS is identical to the RDS:

\bar
\no \frac{\partial \xi_i}{\partial t} &\=& \sjN\,\left\{ \, L_{i,j}\,\xi_j \+ K_{i,j}\,\eta_j\,\right\};\\
 \frac{\partial \eta_i}{\partial t} &\=& \sjN\,\left\{ \, L_{i,j}\,\eta_j \- K_{i,j}\,\xi_j\,\right\}. \label{Jsys}
\ear

Now let us investigate eigenvalues of this system, which will inform about expanding or contracting (unstable or stable) directions. So consider the eigenvalue-eigenvector problem:

\bar
\no  \sjN\,\left\{ \, L_{i,j}\,\xi_j \+ K_{i,j}\,\eta_j\,\right\} &\=& \lambda\,\xi_i;\\
 \sjN\,\left\{ \, L_{i,j}\,\eta_j \- K_{i,j}\,\xi_j\,\right\} &\=& \lambda\,\eta_i. \label{esystem}
\ear

\ni To learn what values of $\lambda$ are possible, it is useful to look at the quantum eigenvalue-eigenvector problem:

\be
\sjN\,H_{i,j}\,\psi_j \= \lambda\,\psi_i,\label{qvalue}
\ee

\ni which, separating real and imaginary parts and substituting $\xi$'s for $Q$'s and $\eta$'s for $P$'s, becomes:

\bar
\no \sjN\,\left\{ \, K_{i,j}\,\xi_j \- L_{i,j}\,\eta_j\,\right\} &\=& \lambda\,\xi_i;\\ 
 \sjN\,\left\{ \, K_{i,j}\,\eta_j \+ L_{i,j}\,\xi_j\,\right\} &\=& \lambda\,\eta_i.\label{qeq}
\ear

\ni Note that (\ref{qeq}) is not identical to (\ref{esystem}). However, if we write in place of (\ref{qvalue}) the perverse equation:

\be
\sjN\,H_{i,j}\,\psi_j \= \srmo\,\lambda\,\psi_i,\label{qvaluep}
\ee
   
\ni and carry out the same procedure, we are lead back to (\ref{esystem}). 

Everyone knows that the quantum problem has all real eigenvalues (which was essential to the \vN\ axioms about measurement). 
We can therefore conclude about the JDS: {\em all eigenvalues are pure imaginary}.
This is not quite right, because in the real context of (\ref{esystem}) but with imaginary eigenvalue, no eigenvector can exist (except possibly for $\lambda = 0$). 
So another way of putting our conclusion about linear quantum mechanics, 
considered as a real dynamical system, is:
{\em there are no expanding or contracting directions}. (Another phraseology would be: {\em there can be no stable and unstable manifolds meeting at any point}.) 
This fact merely reflects the trivial nature of the quantum evolution in the linear case: 
there are in fact $N$ orthogonal eigenvectors in the complex setting, and the dynamics simply multiplies each component of $\psi$ 
in that basis by a phase factor. So each component undergoes a rotation in the complex plane. In other words, finite-dimensional QM produces
Lissajoux figures, but cannot exhibit any more interesting kind of dynamics.

\section{Expanding and Contracting Directions, Nonlinear Case}

I now introduce a nonlinear, finite-dimensional model similar to what was discussed in paper I of this series. For the Hamiltonian I postulate:
\def\HQM{H_{QM}}

\be
\cH \= <\psi|\,\HQM\,|\psi> \+ w\,\left\{\,<\psi|\,S^2\,|\psi> \- <\psi|\,S\,|\psi>^2\,\right\},
\ee

\ni where $\HQM$ denotes the ordinary, linear QM operator and $S$ is some self-adjoint operator. For the evolution equation we need only write:

\be 
\srmo\,\frac{\partial \psi}{\partial t} \= \frac{\partial}{\partial \psi^*}\,\cH.
\ee

\ni (It was observed in paper I and earlier by Weinberg and others that this set-up is merely a repackaging of Hamiltonian mechanics.) 
Next I specialize as follows. For $S$ I assume a diagonal matrix of form

\be
S_{i,j} \= s_i\,\delta_{i,j},\label{diagform}
\ee

\ni with real $s_i$ to be defined later on. For the quantum part I choose the real case: $L = 0$. The dynamical equations become:

\bar
\no \frac{\partial Q_i}{\partial t} &\=& \sjN\,K_{i,j}\,P_j \+ f_i\,P_i;\\
\no \frac{\partial P_i}{\partial t} &\=& \- \sjN\,K_{i,j}\,Q_j \- f_i\,Q_i;\\
f_i &\=& w\,s_i\,\left\{\,s_i \- 2\,\skN\,s_k\,\left(\,Q_k^2 + P_k^2\,\right)\,\right\}.
 \label{QPnlsys}
\ear

\ni This system is nonlinear. The derived Jacobian system I will write as:

\bar
\no \frac{\partial \xi_i}{\partial t} &\=& \sjN\,\left\{\,A_{i,j}\,\xi_j \+ B_{i,j}\,\eta_j\,\right\};\\
 \frac{\partial \eta_i}{\partial t} &\=& \sjN\,\left\{\,C_{i,j}\,\xi_j \+ D_{i,j}\,\eta_j\,\right\},
\ear

\ni where:

\bar
\no A_{i,j} &\=& \- 4\,w\,s_i\,s_j\,P_i\,Q_j;\\
\no B_{i,j} &\=& K_{i,j} \+ f_i\,\delta_{i,j} \- 4\,w\,s_i\,s_j\,P_i\,P_j;\\
\no C_{i,j} &\=& \- K_{i,j} \- f_i\,\delta_{i,j} \+ 4\,w\,s_i\,s_j\,Q_i\,Q_j;\\
 D_{i,j} &\=& \- A_{i,j}
\ear

Consider the corresponding eigenvalue problem:

\bar
\no \sjN\,\left\{\,A_{i,j}\,\xi_j \+ B_{i,j}\,\eta_j\,\right\} &\=& \lambda\,\xi_i;\\
 \sjN\,\left\{\,C_{i,j}\,\xi_j \+ D_{i,j}\,\eta_j\,\right\} &\=& \lambda\,\eta_i. \label{nlvalue}
\ear

I claim this system may have real eigenvalues. To prove the claim I will rely on the following simple theorem:

{\bf Theorem}. Let $M$ be a real, even-dimensional matrix. If 

\be
\det\,M \< 0,
\ee

\ni then the problem 

\be
 M\,X \= \lambda\,X
\ee

\ni has both positive and negative eigenvalues.

{\bf Proof}. Consider the characteristic polynomial:

\be
p(\lambda) \= \det\,\left(\, M - \lambda\,I\,\right).
\ee

\ni Since M is even-dimensional, the leading term is of even order and has coefficient one. Therefore

\be
\lim_{\lambda \to \pm \infty}\,p(\lambda) \= \+ \infty.
\ee

\ni The coefficient of the constant term is $\det M$. Thus, if the latter is negative, the graph of $p(\lambda)$ vs. $\lambda$ must cross the horizontal 
axis in at least one positive and one negative value. QED.
(If $M$ is symmetric corresponding eigenvectors exist; otherwise, I use ``eigenvalue" to mean a solution of $p(\lambda) = 0$.)

To see that the theorem can be relevant, consider the case with $N = 2$ "spins", so $M$ is 4x4 and:

\bar
\no L_{i,j} &\=& 0;\\
\no P_i &\=& 0;\\
\no Q_2 &\=& 0;\\
\no K_{i,j} &\=& k_i\,\delta_{i,j};
\ear

Then an explicit evaluation (left to the reader) yields:

\be 
\det\,M = \left(\,f_2 + k_2\,\right)^2\,\left(\,f_1+ k_1\,\right)\,\left(\,f_1 + k_1 - 4\,w\,s_1^2\,Q_1^2\,\right).
\ee

If $k_1 = - f_1 + \epsilon$:

\be
\det \,M \= \left(\,f_2 + k_2\,\right)^2\,\epsilon\,\left(\,\epsilon - 4\,w\,s_1^2\,Q_1^2\,\right),
\ee

\ni which can be negative if $\epsilon > 0$ and $w$ is large enough.

\section{A Measurement Scenario}

In this section I restrict the model in an attempt to imitate a measurement situation of the type treated in papers I and II for continuum models. I will recur to the traditional setting of
$q$ ``qubits" or ``spins", with spin $J$ = 1/2, meaning the spins take values $\pm 1/2$. Thus the state takes the familiar Dirac form:

\be
\psi \= \sum \, \psi(s_1,s_2,...,s_q)\,|s_1,s_2,...,s_q>.
\ee

There are $2^q$ components, yielding a $2 2^q$-dimensional RDS. The first ``spin" will be singled out as the microscopic system to be measured; the remaining $q - 1$ ``spins" 
will form the ``measuring apparatus",
with readout the ``total spin" of the apparatus components:

\be 
S \= \sum_{k=2}^q\, s_k.\label{Sdef}
\ee

For the quantum part I adopted the usual form:

\be
\HQM \= \- \frac{1}{m}\,\triangle \+ V(S),\label{linearpart}
\ee

\ni where the first term is defined similarly to the continuum case, but using finite differences:

\def\skq{{\sum_{k=1}^q}}
\bar
\no \triangle\,\psi &\=& \skq\,\left\{\,\psi(s_1,...,s_k+1,...,s_q) \+ \psi(s_1,...,s_k-1,...,s_q)\right. \\
 &\-& \left. 2\,\psi(s_1,...,s_k,...,s_q)\,\right\},
\ear

\ni with ``reflecting boundary conditions", meaning $ \psi(s_1,...,s_k+1,...,s_q) =  \psi(s_1,...,s_k,...,s_q)$ if $s_k + 1$ is greater than 1/2, and similarily for subtracting one making a spin argument 
less than - 1/2.

For the external potential I adopted the two-well quartic defined by:

\bar
\no V(x) &\=& \hbox{const.}\,(x - R)^2\,(x + R)^2;\\
\no R &\=& J\,(q - 1);\\
    \hbox{const.} &\=& \hbox{Height}/R^4,
\ear

\ni whose graph for the case: $J = 1/2$ and $q = 9$ is shown in Figure 1. 
\pagebreak

\begin{figure}
\rotatebox{0}{\resizebox{5in}{5in}{\includegraphics{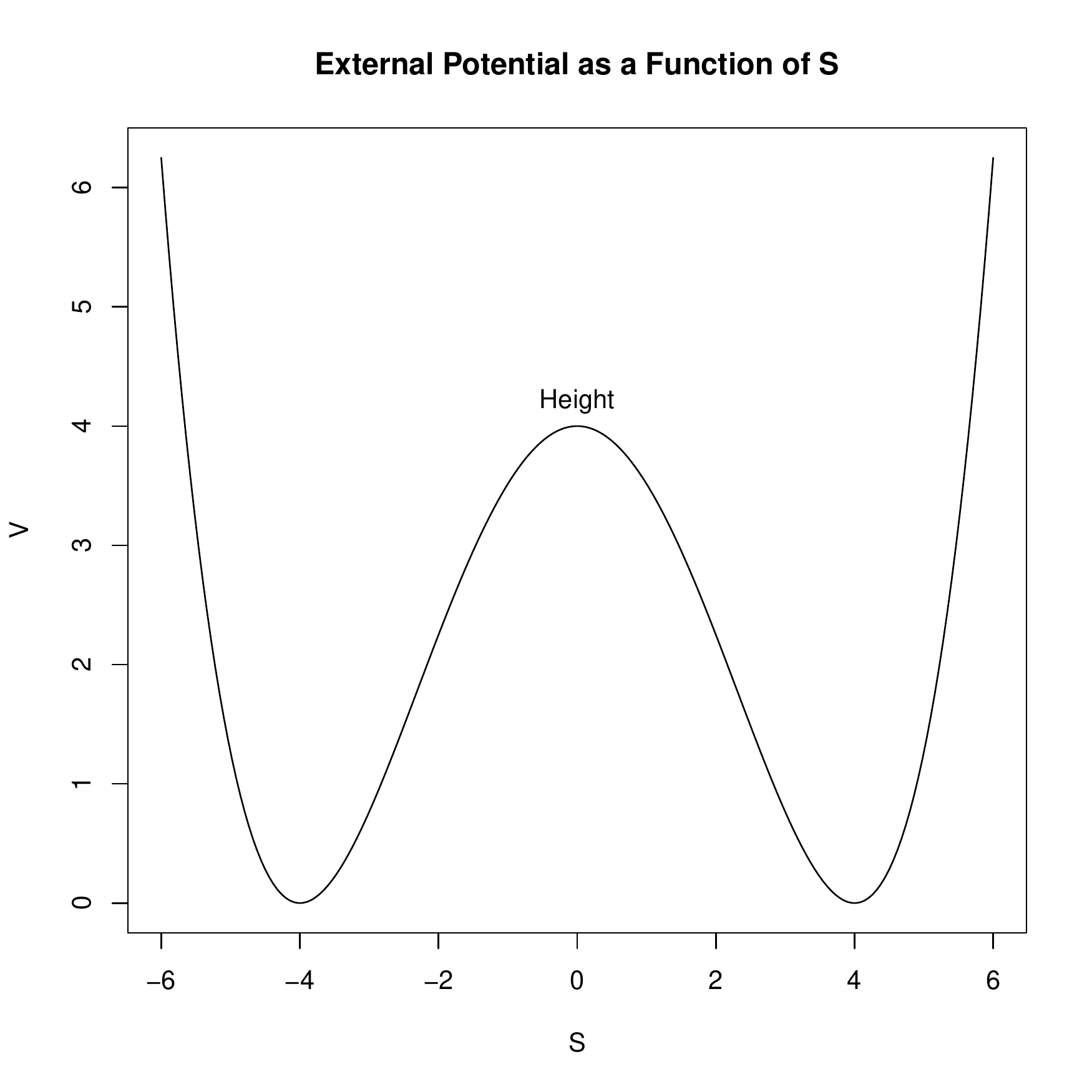}}}
\caption{External potential plotted vs. ``apparatus spin".}\label{fig1}
\end{figure}

To couple the ``total apparatus spin" with the ``microsystem spin" I added to the right side of (\ref{linearpart}) a term:

\be
\alpha\,s_1\,S.
\ee

For the initial state I adopted the superposition:

\bar
\no \psi &\=& z\,\left\{\, \beta_1\,|+ 1/2> + \beta_2\,| - 1/2>\,\right\} \times \\
\no    & & \left\{\,\sum_{s_k = \pm 1/2: |S| \leq \hbox{Center}}\,|s_2>|s_3>...|s_q>\,\right\}.\label{initial}\\
\beta_1^2 &\+& \beta_2^2 \= 1,
\ear
 
\ni where $z$ is a normalization constant and ``Center" is a positive parameter.

The nonlinear terms are the same as in the previous section except that now the diagonal components are ``total apparatus spin", meaning (\ref{diagform}) is replaced by:

\be
S_{i,j} \= S(i)\,\delta_{i,j},\label{diagform1}
\ee

\ni where $S(i)$ denotes the sum appearing in (\ref{Sdef}) corresponding to the component with index $i$.

Table I shows parameters used. I restricted $q$ to odd values so that the ``apparatus total spin" could take the value zero, centering the simulations. 

\vskip0.1in
\centerline{Table I. Parameter Values}
\vskip0.1in
\begin{center}
\begin{tabular}{|l|l|}
\hline
$q$ & 5,7, or 9  \\ \hline
Height & 1.0 or 10.0 \\ \hline
Center & 3.0 \\ \hline
$w$ & 0.0 or 2.2  \\ \hline
$\beta_1/\beta_2$ & 1.0 or 1.2 \\ \hline
Inverse mass & 0.1 \\ \hline
$\alpha$ & 1.0 \\ \hline
\end{tabular}
\end{center}

I first examined the determinant of the JDS for the symmetric superposition ($\beta_1 = \beta_2$) at the initial state. 
For the linear case ($w = 0$) $\det M$ was positive, as we know it must be from results in Section two. 
For the nonlinear case, as $w$ was increased in steps of 0.1, a
threshold appeared where the determinant turned negative. The threshold showed a trend: increasing with the steepness of the external potential, and decreasing with increased $q$.  
The asymmetrical cases ($\beta_1/\beta_2 = 1.2$) produced the same result. See Table II. 
\pagebreak

\vskip0.1in
\centerline{Table II. Threshold Values of ``$w$"}
\vskip0.1in
\begin{center}

\begin{tabular}{|l|l|l|}
\hline
q & Height & $w$(thres.) \\ \hline
5 & 1.0 & 0.55 \\
7 &     & 0.45 \\
9  &   &  0.45 \\ \hline
5 & 10.0 & 2.35 \\
7 &    & 2.15 \\
9 &   & 2.05 \\ \hline
\end{tabular}
\end{center}

I next made simulations for $q=9$ in four cases. Figure 2 shows the density of $\psi$ as a function of ``apparatus total spin" for the 
symmetric linear case ($w = 0$) at ``time 10.0";  wavepackets moved to right and left, forming a cat. Figure 3 shows the same except for making $w$ positive;
cats can't form, but with no force to break the symmetry nothing happens. Figure 4 shows the asymmetric, $w = 0$ case; one gets asymmetric cats. 
Figure 5 is the interesting case: asymmetric with positive $w$; the needle on the register moved to the left. 

\begin{figure}
\rotatebox{0}{\resizebox{5in}{5in}{\includegraphics{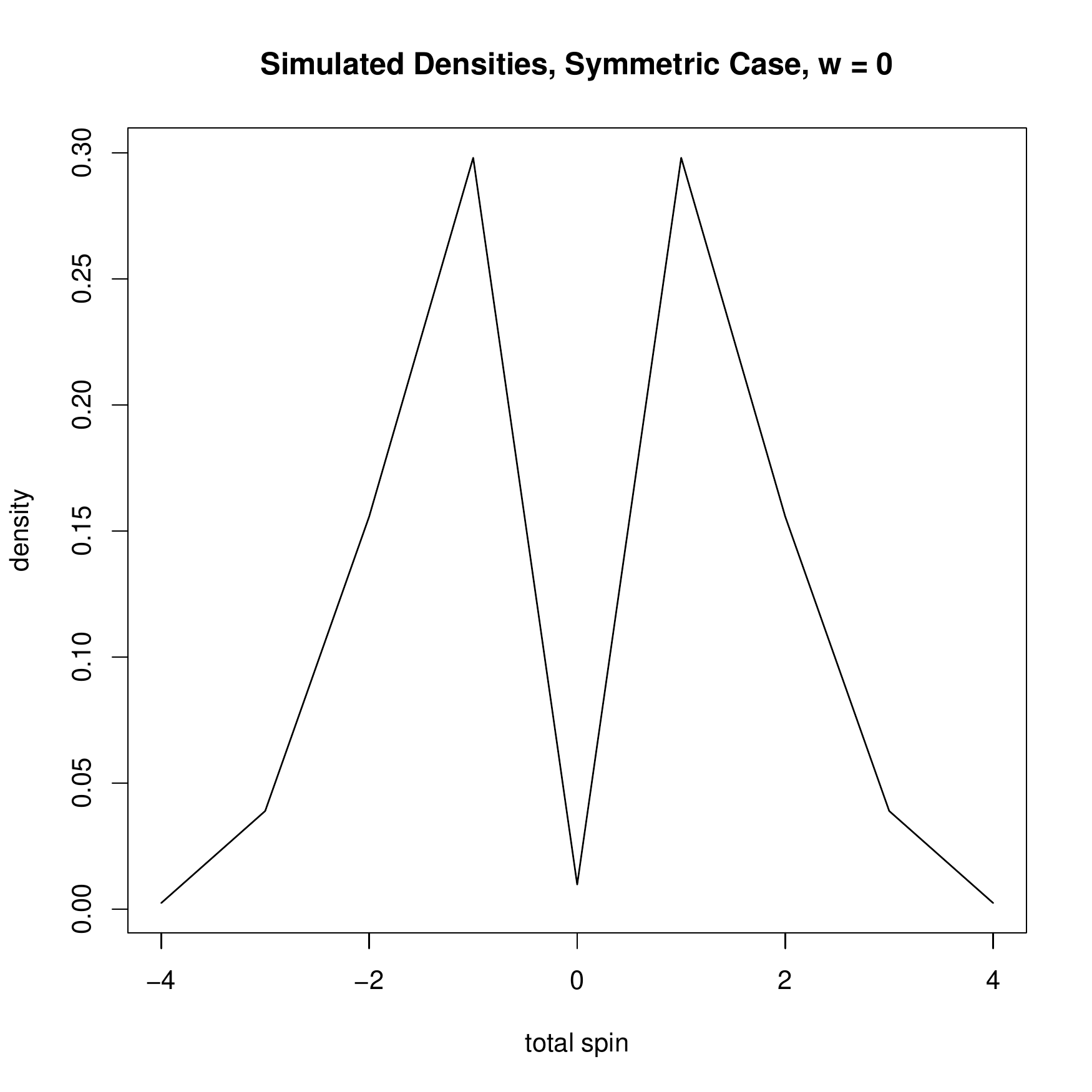}}}
\caption{Density vs. total spin for the linear, symmetric case.}\label{fig2}
\end{figure}

\begin{figure}
\rotatebox{0}{\resizebox{5in}{5in}{\includegraphics{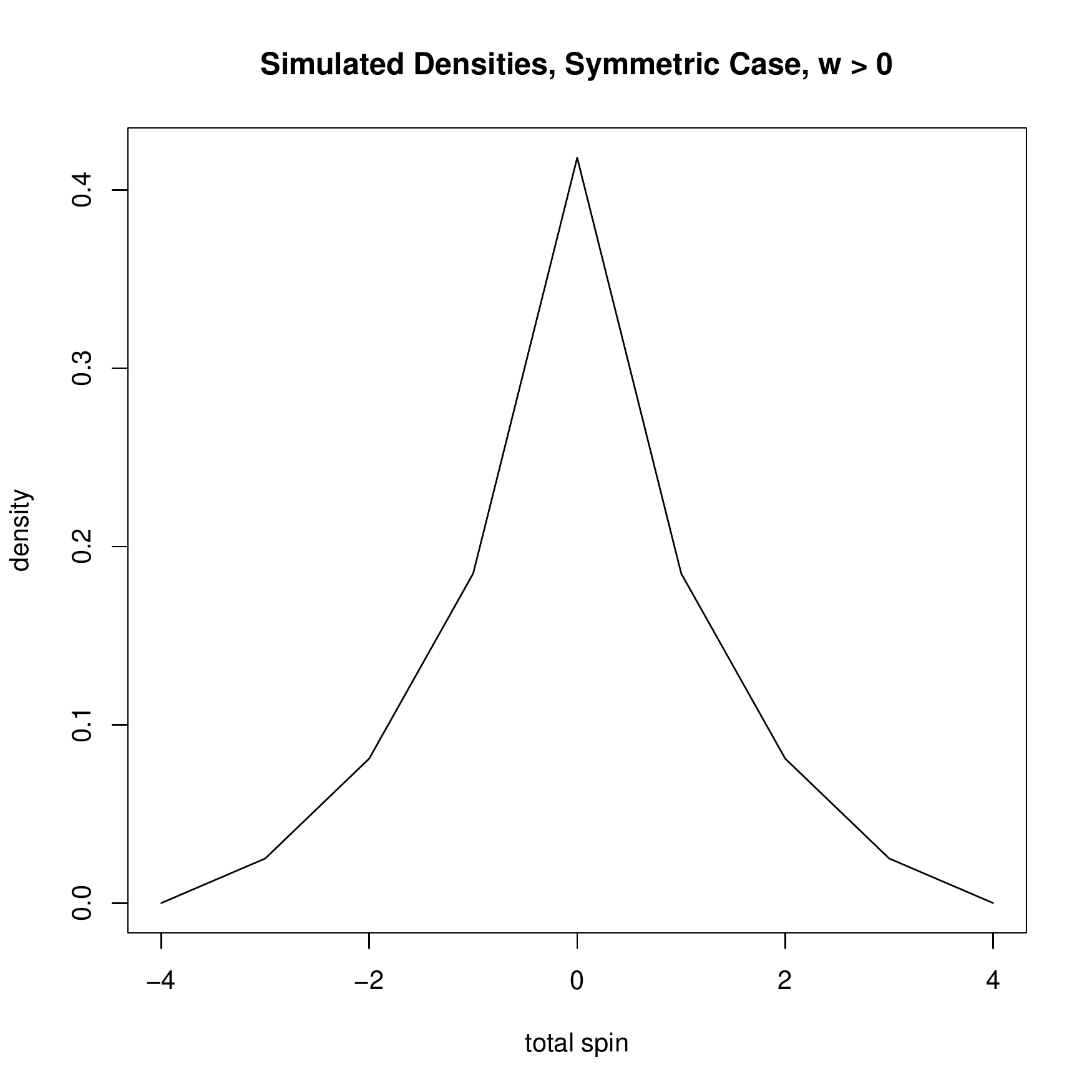}}}
\caption{Density vs. total spin for the nonlinear, symmetric case.}\label{fig3}
\end{figure}

\begin{figure}
\rotatebox{0}{\resizebox{5in}{5in}{\includegraphics{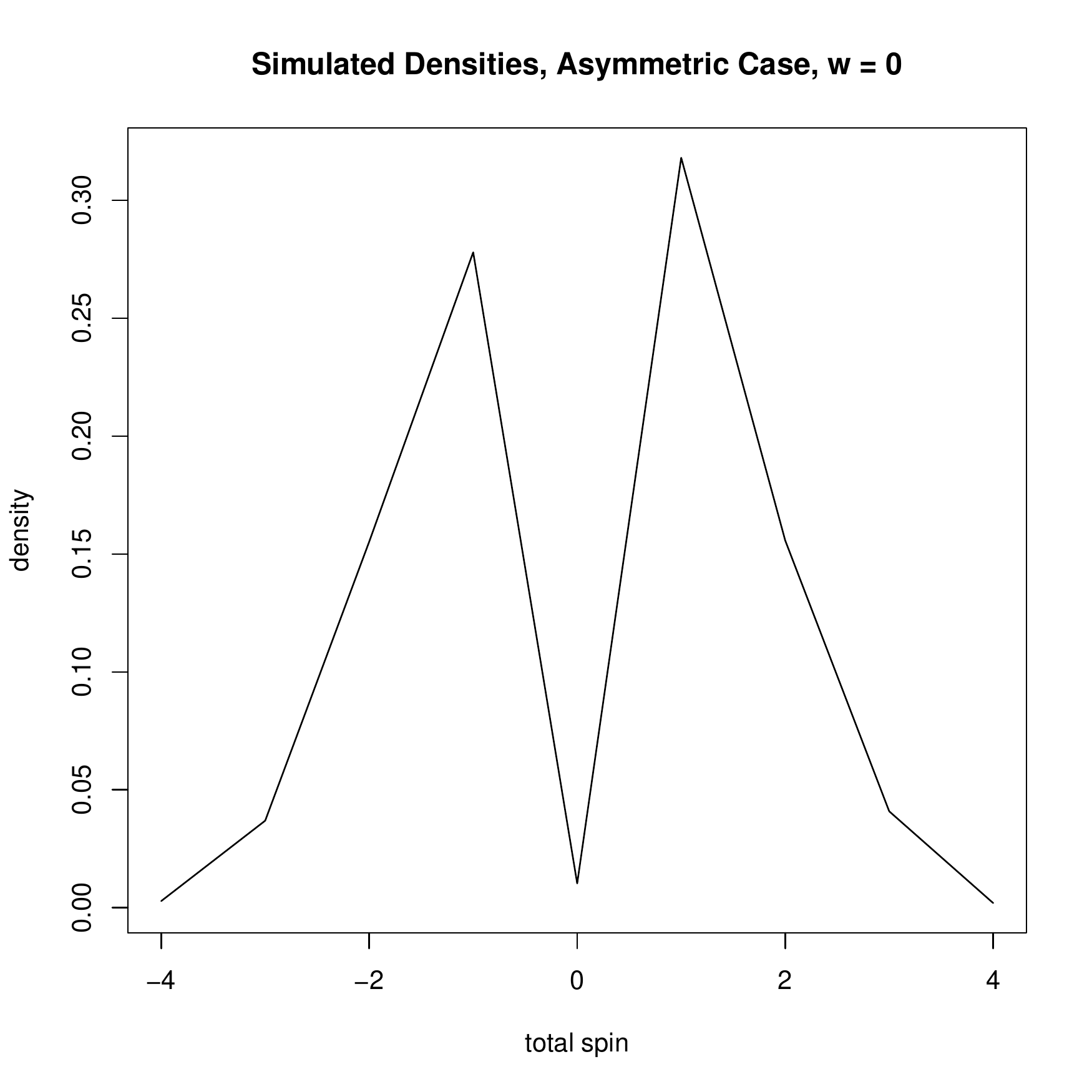}}}
\caption{Density vs. total spin for the linear, asymmetric case.}\label{fig4}
\end{figure}

\begin{figure}
\rotatebox{0}{\resizebox{5in}{5in}{\includegraphics{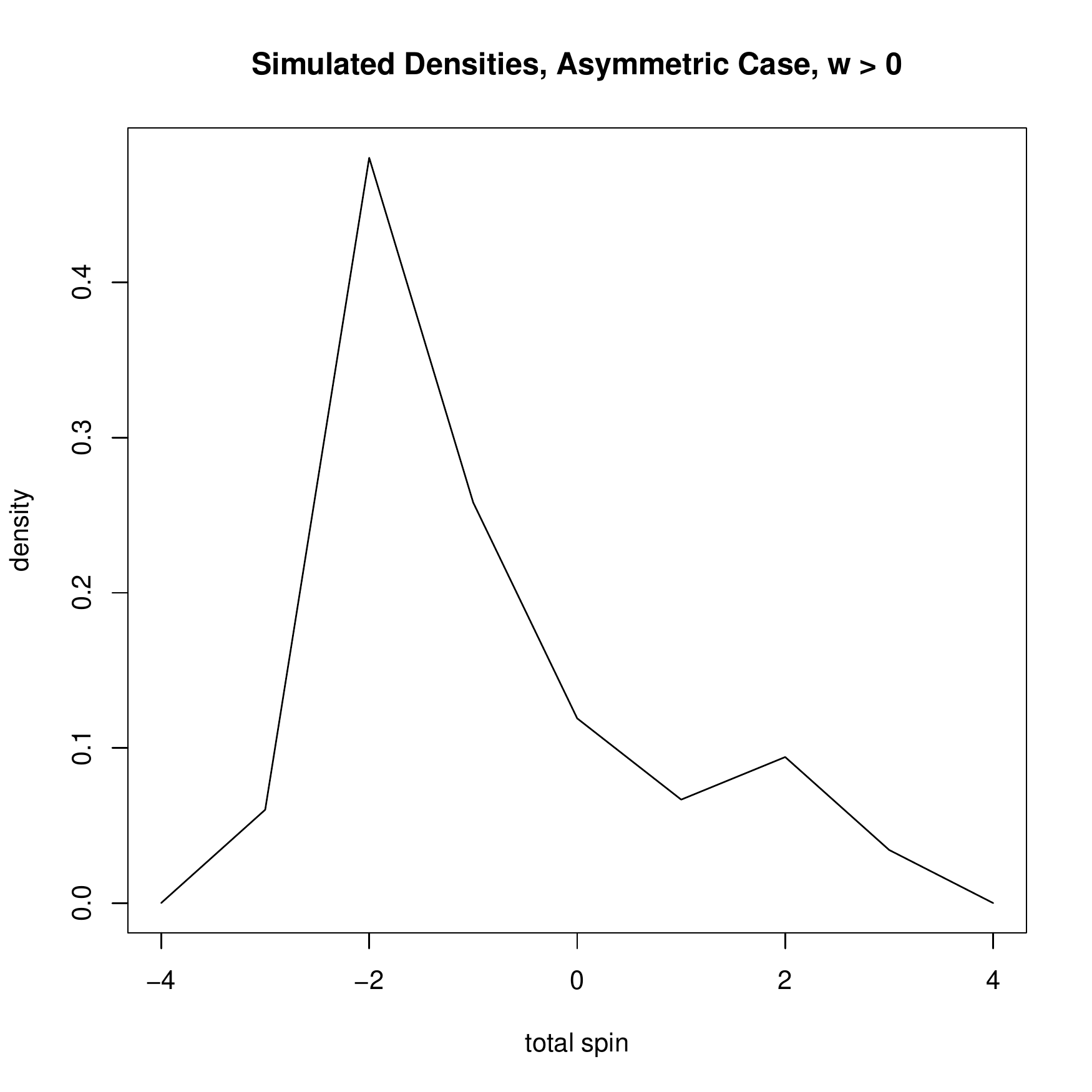}}}
\caption{Density vs. total spin for the nonlinear, asymmetric case.}\label{fig5}
\end{figure}
\pagebreak

In order to evaluate the dependence on initial conditions, I randomized the initial wavefunction components 
and studied registrations, which somewhat arbitrarily I defined as 50 percent higher
total density for apparatus spin on the left vs. the right and {\em vice versa}. Thus three outcomes were possible: right registration, 
left registration and ``no detection", meaning no movement of the needle, or cat formation. For one kind of randomization, call it component-wise, I replaced $Q_i$ by 
$Q_i\,(1 + r_i)$ or by $r_i$, depending on whether $Q_i$ is nonzero or zero in the initialization, equations (\ref{initial}), and $P_i$ by $r^{'}_i$, and then normalized the wavefunction.
The $r_i$, $r^{'}_i$ were taken i.i.d. and 
either $\sigma$ times a uniform random variable on the interval $(-1,1)$, or $\sigma$ times a standard normal random variable (mean zero and variance one).
Some results are shown in Table III. 
Abbreviations: Reps, repetitions; Rand, choice of randomization, uniform or normal;
 ``LRs", left registrations; ``RRs", right registrations; ``NDs", no detections;  ``BP", usual Born/\vN\, probability of ``finding the spin up", $\beta_2^2$; 
``SP": probability computed from simulations ignoring NDs: RRs/(RRs + LRs). Nonlinear case, $w$ = 2.2, parameter ``Height" = 10.0, and $q = 9$.
 
\vskip0.1in
\centerline{Table III. Registrations and NDs, Component-Wise Randomization}
\vskip0.1in
\begin{center}

\begin{tabular}{|l|l|l|l|l|l|l|l|}
\hline
Reps & Rand & $\sigma$ & LRs & RRs & NDs & BP & SP \\ \hline
100 & unif.& 0.1 & 30  & 32  & 38  & 0.5  & 0.516 \\ \hline
100 & norm.& 0.1 &44  & 34  &  22  & 0.5  & 0.436 \\ \hline
50 & unif. & 0.1 &46  & 0  & 4  & 0.4 & 0.0  \\ \hline
50 & norm.  & 0.1 &39 &  8 & 3 & 0.4 & 0.17  \\ \hline
50& unif.  & 0.1 &49 & 1 & 0  &  0.3 &  0.02 \\ \hline
50 & norm.  & 0.05 & 50 & 0 & 0 & 0.4 & 0.0  \\ \hline
\end{tabular}
\end{center}

I also implemented an idea suggested in paper II and perhaps traceable to Deutch (see citation there): 
a random unitary transformation of the wavefunction. Accordingly I generated random $N \times N$ unitary matrices by
setting:

\def\Kr{{K_{\hbox{rand.}}}}
\be
U \= \exp\,\left\{\,\sqrt{-1}\,\delta\,\Kr\,\right\},
\ee

\ni where $\Kr$ denotes a symmetric real matrix filled with i.i.d. standard normals and $\delta$ is a small ``time" interval. 
($U$ was generated at the start of each run by use of the symplectic solver, see Computational Appendix.)
Some results are shown in Table IV.

\vskip0.1in
\centerline{Table IV. Registrations and NDs, Unitary Randomization}
\vskip0.1in
\begin{center}

\begin{tabular}{|l|l|l|l|l|l|l|}
\hline
Reps & $\delta$ & LRs & RRs & NDs & BP & SP \\ \hline
50 &  0.02  & 18  & 18  & 14  & 0.5 & 0.5 \\ \hline
50 &  0.02  & 14  & 9  & 14  & 0.4 & 0.25 \\ \hline
50 &  0.05  & 0  & 0  & 50  & 0.4 &  - \\ \hline
\end{tabular}
\end{center}

\pagebreak
\pagebreak

\section{Discussion}

Both ``chaos" and ``cat" are somewhat ambiguous terms. Starting with the former, consider the famous claim that the 
the outer planets of the solar system---Jupiter, Saturn, Uranus, and Neptune, ignoring the since-demoted Pluto---are chaotic. In a paper in Science in 1999, \cite{sciencepaper}, N. Murray and M. Holman
reported, based on some analysis and some simulations, that, making a 1.5 mm shift in the initial position of Uranus, the orbits subsequently diverged. 
But time-scale is an issue here---they mention a few million years. 
If the scale of divergence instead were a few billion years, would we still agree that this planetary deviance represents ``chaos"? 
As for ``cats", the reader surely noted my adoption of an arbitrary criterion for cats vs. no-cats. However, in the context of a realistic measurement, 
knowing the apparatus and system parameters will eliminate
the ambiguities. The time scale is simply that of each ``run", while ``cat-or-no-cat" is a matter of device sensitivities.
(In the simulations, the time-scale was chosen to see a density peak moving to right or left and the parameter ``Height" controlled the sensitivity.)
We can summarize the philosophy of measurement supported here in the aphorism: ``The apparatus makes the observable."

The study of the sign of the Jacobian determinant also contained an ambiguity: I arbitrarily declared that ``$\det\,M <0$" if it was less than -0.01. 
Although I have shown a qualitative and quantitative difference between linear QM and my nonlinear version, perhaps the magnitude of the real eigenvalues 
in the latter would be more informative than determinants. 
Conceivably, the strength of randomness required to produce frequent left- or right-detections might be determined by these eigenvalues. 
Since the eigenvalues have units of energy, it is tempting to compare them with the expected energy perturbation
caused by the random part of the wavefunction (where by ``expected" I refer to average over the random part, and NOT to the meaning in \Copism). However, contemplating 
the continuum model discussed in papers I and II, one would conjecture that the relevant quantities are the height of the external potential, 
the coupling of microsystem with apparatus,
the size of the ``kick" given through the random part, and the time-scale of the experiment (a kick of any size 
should generate motion to right or left, but if too small the needle may not move during the ``run"). Perhaps that macroscopic perspective is overly simplistic, 
because it assumes free interchange of integrals over
$V^{'}(S)$ with that function applied to the integrated total spin (not the case in this work). 
As the apparatus sensitivity is increased, 
eventually the dispersion of the system will not be smaller than the scale on which $V$ varies. In that case, this classical picture might break down, at least when the decision (right or left?) is being made. 

At this time, I cannot propose a theorem combining these quantities---physical characteristics of the apparatus and eigenvalues/eigenvectors of the Jacobian matrix---into a comprehensive explanation of the
randomness required to explain measurement outcomes. Thus, for the studies, I experimented to find variances that at least affected the outcomes.
   
I also cannot prove that the appearance
of expanding and contracting directions in the linearized (Jacobian) approximation means that the full nonlinear system exhibits a coincidence of stable and unstable manifolds, implying
a separatrix exists locally between distinct behaviors, so that a small perturbation might have a large eventual impact. (The ``strong chaos" in the solar system  Murray and Holman 
attributed to resonances between nearby periodic orbits, a scenario that 
probably isn't relevant in this context.) 

How much of \Poin's conception of probability have I demonstrated in this work? 
 A clear distinction appeared between the symmetric (microsystem ``up" and ``down" have equal weights) and asymmetric cases.
In the former, different randomization schemes lead to the same result. But in the latter, they clearly deviated. This might be expected. Consider again roulette, and imagine that the wheel
has higher probabilities of capturing the ball in the first half of numbers than for the second half (say, by making the first-half of slots stickier than are the others). 
Then the croupier might well be able to cheat the
player. That is, bias in the croupier's toss now becomes relevant, where in the symmetrical case it would not matter. 

I can say something more concrete for the particular set-ups used in this work. First: the component-wise randomization of the initial wavefunction. 
Consider the force on the apparatus (see paper II for why this is the relevant quantity):

\be
F \= \alpha\,<\psi|\,s_1\,|\psi>,
\ee

\ni which takes the form:

\be
F \= \alpha\,\left(\,\beta_1^2\,X_{-} \- \beta_2^2\,X_{+}\,\right).
\ee

\ni Here $X_{-}$ and $X_{+}$ are i.i.d. random variables of the type:

\bar
\no X_{\pm} &\=& \frac{1}{n}\,\sum_{i: |S(i)| < C}1\left[\,s_1(i) = \pm J\,\right]\,\left\{\,1 + r_{\pm}(i)\,\right\}^2;\\
    n    &\=& \sum_{i: |S(i)| \leq C}\left\{\,\beta_1^2\,\left(\,1 + r_{+}(i)\,\right)^2 + \beta_2^2\,\left(\,1 + r_{-}(i)\,\right)^2\,\right\} .\label{nequ}
\ear

\ni where $1[\,\cdot\,]$ denotes indicator function
and the $r_k(i)$ are i.i.d. with some selected distributions, e.g., uniform or normal. 
As in paper II we can postulate that there is a threshold, call it $\tau$, for which:

\bar
\no P\left[\,\hbox{LR}\,\right] &\=& P\left[\, F < - \tau\,\right];\\
P\left[\,\hbox{RR}\,\right] &\=& P\left[\, F > \tau\,\right]
\ear

\ni (If such a ``hard" threshold exists, it will be a function of all the device parameters mentioned in a previous paragraph.)

Various possibilities arise. In the case: $\beta_1 = \beta_2$, $F$ has a symmetrical distribution with mean zero, so whatever are the distributions of the $r_k(i)$, we will find
$P\left[\,\hbox{LR}\,\right] = P\left[\,\hbox{RR}\, \right]$, and the variances simply fix how many NDs occur relative to detections. 
In the asymmetrical case, the crucial issues are the variances of the $r_k(i)$'s, the number $n$ in the normalization, equation (\ref{nequ}), which is up to a constant approximately 
the number of nonzero initial wavefunction components,
and the ratio: $\tau/\alpha$. Write $F = \alpha\,\theta$. By adding and subtracting a constant we can write

\be
\theta \= 2\,m\,\left(\,\beta_1^2 - \frac{1}{2}\,\right) \+ \tilde{\theta},
\ee

\ni where $\tilde{\theta}$ has mean zero and, by the Central Limit Theorem if $n$ is large, $m \approx 1/2$ and

\be
\tilde{\theta} \= \hbox{O}\left(\frac{\sigma}{\sqrt{n}}\,\right).
\ee

So we can write, e.g.:

\be
P\left[\,\hbox{RR}\,\right] \= P\left[\,\left(\,\beta_1^2 - \frac{1}{2}\,\right) > \tilde{\theta} + \frac{\tau}{\alpha} \right].\label{peq}
\ee

In the unitary-randomization case, provided $\delta$ is small enough to justify an expansion, one finds an expression:

\bar
\no \theta &\=& 2\,\left(\,\beta_1^2 - \frac{1}{2}\,\right) \+ \delta^2\,\sum_{i,j,k\, \hbox{rest.}}\,1\left[\,s_1(i) = + J\,\right]\,K_{i,j}\,K_{j,k} \- \\ 
&& \delta^2\,\sum_{i,j,k\, \hbox{rest.}}\,1\left[\,s_1(i) = - J\,\right]\,K_{i,j}\,K_{j,k},
\ear

\ni where ``rest." means restriction by: $S(\hbox{index}) \leq \hbox{Center}$. 
(For this to hold, clearly $\delta$ must be small enough that typical values of the sums over $K$'s multiplied by $\delta^2$ are small.)
\vskip0.1in

Suppose $\tau/\alpha \approx 0$. Then, depending on variances and $n$, we can get cases:

If $\sigma$ or $\delta$ is too small, the random perturbation may be insufficient to induce perceptible motion in the pointer (presumably because the wavefunction remains on the original side of a separatrix 
between distinct eventual macroscopic displacements).   

If $n$ is large, in the component-wise case a measurement may only be able to distinguish $\beta_1^2 = 1/2$ from $\beta_1^2 \neq 1/2$ (dichotomous outcome). 
In the unitary case, the effect of the randomization may be to circularize the
initial state, resulting in large numbers of, or exclusively, no detections.

If $n$ is not large, the above probability will depend on $\beta_1^2$. Various laws can arise, for example, if $\tilde{\theta}$ is uniformly distributed on the interval
$(-1/2,1/2)$ then the probability in (\ref{peq}) equals $\beta_1^2$, that is, Born's law holds. However, many other laws can appear, and Born's law is not singled out by the set-up.

If Born's axiom doesn't hold,
it could be said the situation does not represent a \vN\ measurement. (Then again,
the same can be said of the original Stern-Gerlach experiment, as pointed out in paper II.)
This is only worrisome 
if you accept the Copenhagen interpretation. On the other hand,
it could be the basis of falsifying the present theory experimentally.

In the measurement scenario of section 4, if the simulations were run long enough, it is conceivable that the needle shown in Figure 5 that moved left might carrom off the wall
and move back to the right! (Because of the reflecting boundary conditions adopted in the definition of the kinetic-energy operator.) 
Alternatively, the needle might oscillate in the left potential well, as
was seen in paper II assuming classical behavior of the apparatus. 
Running a simulation with parameters as for Figure 5 for quadruple the time, the rebound appeared, see Figure 6.
Clearly, this phenomena is a consequence of the discrete spins resulting in hard walls at $S \= \pm (q-1)\,J$ and the 
small size of the macrosystem, and is probably irrelevant in the continuum case with realistic apparatus. 

\begin{figure}
\rotatebox{0}{\resizebox{5in}{5in}{\includegraphics{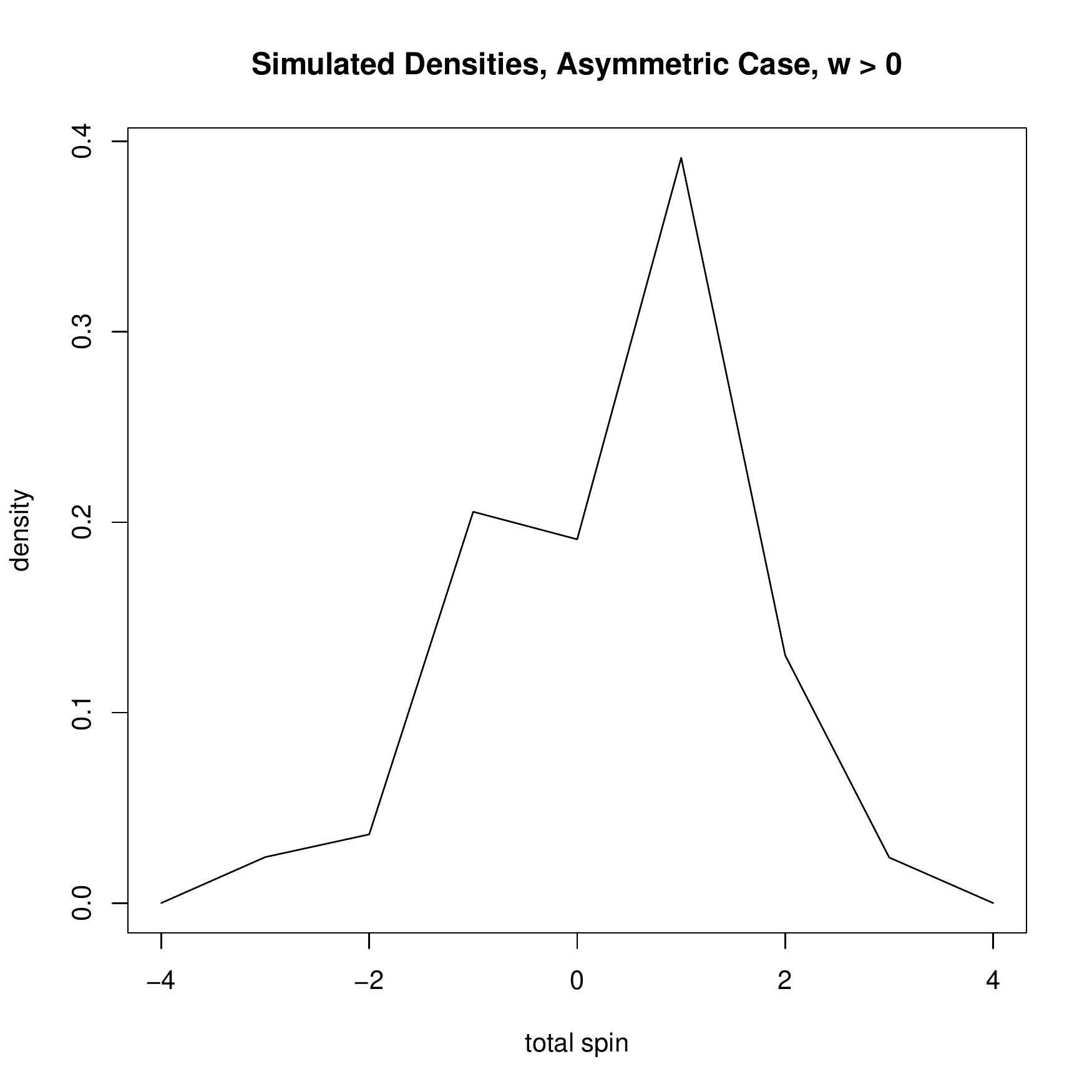}}}
\caption{Density vs. total spin for the nonlinear, asymmetric case, longer time.}\label{fig6}
\end{figure}
\pagebreak

This paper provides some (weak) evidence for a measurement scenario in which a large (many-component) apparatus coupled to a microsystem is able to generate apparent randomness,
because a small variation in initial conditions is amplified up to observable levels. 
The results should be reproduced for larger systems with more realistic parameters (in particular, a smaller coupling constant). 
Perhaps both issues will be addressed when faster (quantum?) computers become available. 
Alternatively, 
a theory of the ultimate source of randomness might be enlightening.

\section{Computational Appendix}

To elucidate the issues affecting computer simulations of nonlinear Hamiltonian systems, I have to explain the concepts: ``symplectic", ``solver order", ``explicit vs. implicit", 
and ``separable vs. nonseparable Hamiltonians".

To get started: the time-evolution of finite-dimensional, linear or nonlinear, QM is given by an ODE system. However, one is advised against solving the system using the methods (such as Runge-Kutta) 
provided in standard mathematical packages, because these methods do not preserve the ``symplectic structure" of Hamiltonian mechanics, the volume element, or the energy. 
Indeed, trying to solve Kepler's model on these packages, one can watch as the orbit shrinks. Hence, one must acquire a so-called ``symplectic solver". Many such recipes have been described; 
they fall into two classes called ``explicit" and ``implicit".``Explicit" means that at each time-step one need only evaluate some formulas whose coefficients are determined in previous steps. 
``Implicit" means that you have to solve some equations at each step.``Solver order" refers to the error incurred in the approximation, which is usually indicated in the form: O($h^n$),
$h$ (NOT Planck's constant) being the traditional symbol in numerical analysis for the time-step, and $n$ is the order. 
The minimum useful order is two, meaning per-step error of O($h^2$), because the time-step
is usually related to a preselected final time $T$ of the simulation by: $h = T/\hbox{Steps}$; so, with order one, making that many steps could incur an overall error of order one. 
Mostly one prefers a solver of order three or higher. Finally, Hamiltonians are either ``separable" or ``nonseparable". Separable means that $\cH$ can be written as a sum of terms depending only on Q's
and a sum of terms depending only on P's. E.g., ordinary linear quantum mechanics with real kinetic energy and potentials (so in this paper, $L = 0$) is separable. ``Nonseparable" means there
are terms involving products of Q's and P's.

For this work, I needed to simulate a high-dimensional, nonlinear system, which is moreover nonseparable. (The nonseparability problem derives from the centering term in the nonlinear energy.)
To do so I needed an explicit method, because solving a high-dimensional nonlinear equation system at each time step 
would tremendously slow down the simulations, while moreover raising many issues about multiple solutions, convergence of Newton-Rafson or whatever function-maximizer or zero-locator is used, etc. 
Over the years, most authors writing on this subject maintained that explicit methods for nonseparable Hamiltonians do not exist, although some had been exhibited in special cases (for references,
see \cite{tao}). 

However, in 2016 M. Tao published a ``generic" solver for nonseparable Hamiltonians, \cite{tao}. 
The method was based on the following idea: introduce variables $X$ and $Y$ ``doubling" $Q$ and $P$,
and an extended Hamiltonian:

\bar
\no \cH_{\hbox{ext.}} &\=& \cH_A(Q,Y) \+ \cH_B(X,P) \+ \cH_C(Q,P,X,Y);\\
 \cH_C &\=& \omega\,\left\{\,||\,Q - X\,||^2 \+ ||\,P - Y\,||^2\,\right\},
\ear

\ni where $\cH_A$ and $\cH_B$ are copies of the original Hamiltonian $\cH$, and $\omega$ is a coupling constant. 
Assuming the original energy function is positive, which is the case for my application,
and $\omega$ chosen large, energy conservation will force approximate agreement of $Q$ with $X$ and $P$ with $Y$. Then uniqueness of solutions of ODEs
in turn forces each pair, $(Q,P)$ and $(X,Y)$ to be approximate solutions of the original system. 
The method evades the nonseparability of the original Hamiltonian by the trick of updating the pair $(Q,Y)$ using derivatives of $\cH_A(X,P)$ and $(X,P)$ using derivatives of $\cH_A(Q,Y)$, 
call these ``update A" and ``update B",
while updating all four variables using $\cH_C$ exactly (as that system is a simple harmonic motion), call it ``update C". Tao then updates in the order: A, B, each by a half-step ($h/2$), 
then C by a full step ($h$),
then B, then A again by half-steps. The claim is made that this procedure yields per-step error of $\hbox{O}(h^3)$. (Tao describes higher-order methods but I did not pursue them.)

I also faced the difficulty that, unable to produce an exact solution of the nonlinear system, I had no procedure to check for programming errors. I therefore pursued a strategy of
comparing two simulation methods. I programmed a routine introduced by Ruth in 1983 for separable systems, \cite{ruth}, which is symplectic, explicit, and fourth-order. 
I evaluated it on the usual test case (the Kepler problem), for which it preserved energy, angular momentum, area, and agreed with the explicit solution 
to at least 5 decimals (for a suitably small time-step). 
Next I applied Ruth's method to the linear QM case, with matrix $K$ described in Section three, and noted that it preserved energy and norm to
that number of places (using 10,000 steps). In the nonlinear case, the nonlinearity and the nonseparability enter through the vector called $f$ in equations (\ref{QPsys}). Ruth's method
has four stages in which one updates P's or Q's with some cleverly selected coefficients. So I kept $f$ constant in each stage but updated it before the next stage. 
This trick yields a symplectic solver (because each of Ruth's stages represents a symplectic, also called ``canonical", update), but it is not guaranteed to be high-order or to preserve energy.
Indeed, since an error in the derivatives of $\cH$ occurs of order $h$ during the time-step, probably the best we can hope for is O($h^2$), the minimal order. 

I then programmed Tao's method. As Tao remarks, 

\begin{quote}
One may worry that large $\omega$ requires a small [time step], which would undermine the computational efficiency gained by an explicit integrator.
\end{quote}

\ni Indeed, testing my implementation of Tao's method on the Kepler problem, I noted that it was necessary to adopt $\omega \approx 10^4$ and 10,000 time-steps 
in order to get the orbits to close off, making an ellipse
as opposed to a Lissajoux figure with incommensurable periods. However, for the NLQM problem Tao's method was the clear winner. Both methods preserved the norm of $\psi$.
But the energy using my trick varied by one to ten percent over the run, depending on the parameters. Tao's method (same $\omega = 10^4$ and time-steps as above) preserved energy to three decimals.
The figures were qualitatively similar. This suggests there were no implementation bugs, but of course there might still be one that affects both methods as they shared certain routines.

Some other computing issues: the determinants were calculated using the LU decomposition as described on p.45 of the book {\em Numerical Recipes in C}, \cite{nmc}. 
However, with the high number of dimensions
for the cases of $q$ = 7 or 9, overflow became a problem. So I conditioned the matrix $M$ by premultiplying all components by a factor of 1/5 or 1/10 respectively, 
then postmultiplied by the appropriate
factor after running the algorithm. 

The program was written in the C language and computations were run on the author's ten-year-old PC. 
(A single simulation using Tao's method required about 20 minutes, so making Tables III and IV required one or two days per line.)
The ambitious reader who wishes to check or extend the results of this paper should do so on a modern platform (a supercomputer, say, or in the cloud) 
in a modern programming style.


\begin{thebibliography}{9}

\bibitem{wick}
Wick, W. D. ``On Non-Linear Quantum Mechanics and the Measurement Problem I: Blocking Cats", Arxiv 1710.03278, and ``II: The Random Part of the Wavefunction", Arxiv 1710.03800, published October 2017.  

\bibitem{poin}
\Poin, H. Calcul des Probabilit{\' e}. Paris (1896).

\bibitem{born}
Born, M. ``Zur Quantummechanik der Stossvorg{\" a}nge". Zeitschrift f{\"u}r Physik. 37: 863-67, (1926). Translated and reproduced in: Quantum Theory and Measurement, 
J. A. Wheeler and W. H. Zurek, Eds., Princeton University Press, Princeton, NJ, 1983.


\bibitem{vnbook}
Von Neumann, J. {\em Mathematical Foundations of Quantum Mechanics}, English translation of the original German text of 1932,
Princeton University Press, Princeton, NJ. (1955).

\bibitem{sciencepaper}
Murray, N. and Holman, M. ``The Origin of Chaos in the Outer Solar System''. Science 283: 1877-81 (1999). 


\bibitem{ruth}
Ruth, R. D. ``A canonical integration technique". IEEE Trans. on Nuclear Science NS30 (4) 2669-2671. (1983).

\bibitem{tao}
Tao, M. ``Explicit symplectic approximation of nonseparable Hamiltonians: algorithm and long-time performance." Arxiv 1609.02212v1, September 2016; Phys Rev E. 94: 043303 2016.


\bibitem{nmc}
Press, W.H. \ea. {\em Numerical Recipes in C}. Cambridge University Press, Cambridge, UK. (1988).
 
\end{thebibliography}
\end{document}